\documentclass{article}
\usepackage{amsmath, amssymb, amsfonts}
\title{ Regular Schwarzschild-like spacetime embedded in a five dimensional bulk} 
\author{Hristu Culetu, \\Ovidius University, Dept.of Physics, \\B-dul Mamaia 124, 900527 Constanta, Romania \footnote{electronic address: hculetu@yahoo.com}}

\begin{document}
\numberwithin{equation}{section}
\pagenumbering{arabic}
\maketitle
\newcommand{\fv}{\boldsymbol{f}}
\newcommand{\tv}{\boldsymbol{t}}
\newcommand{\gv}{\boldsymbol{g}}
\newcommand{\OV}{\boldsymbol{O}}
\newcommand{\wv}{\boldsymbol{w}}
\newcommand{\WV}{\boldsymbol{W}}
\newcommand{\NV}{\boldsymbol{N}}
\newcommand{\hv}{\boldsymbol{h}}
\newcommand{\yv}{\boldsymbol{y}}
\newcommand{\RE}{\textrm{Re}}
\newcommand{\IM}{\textrm{Im}}
\newcommand{\rot}{\textrm{rot}}
\newcommand{\dv}{\boldsymbol{d}}
\newcommand{\grad}{\textrm{grad}}
\newcommand{\Tr}{\textrm{Tr}}
\newcommand{\ua}{\uparrow}
\newcommand{\da}{\downarrow}
\newcommand{\ct}{\textrm{const}}
\newcommand{\xv}{\boldsymbol{x}}
\newcommand{\mv}{\boldsymbol{m}}
\newcommand{\rv}{\boldsymbol{r}}
\newcommand{\kv}{\boldsymbol{k}}
\newcommand{\VE}{\boldsymbol{V}}
\newcommand{\sv}{\boldsymbol{s}}
\newcommand{\RV}{\boldsymbol{R}}
\newcommand{\pv}{\boldsymbol{p}}
\newcommand{\PV}{\boldsymbol{P}}
\newcommand{\EV}{\boldsymbol{E}}
\newcommand{\DV}{\boldsymbol{D}}
\newcommand{\BV}{\boldsymbol{B}}
\newcommand{\HV}{\boldsymbol{H}}
\newcommand{\MV}{\boldsymbol{M}}
\newcommand{\be}{\begin{equation}}
\newcommand{\ee}{\end{equation}}
\newcommand{\ba}{\begin{eqnarray}}
\newcommand{\ea}{\end{eqnarray}}
\newcommand{\bq}{\begin{eqnarray*}}
\newcommand{\eq}{\end{eqnarray*}}
\newcommand{\pa}{\partial}
\newcommand{\f}{\frac}
\newcommand{\FV}{\boldsymbol{F}}
\newcommand{\ve}{\boldsymbol{v}}
\newcommand{\AV}{\boldsymbol{A}}
\newcommand{\jv}{\boldsymbol{j}}
\newcommand{\LV}{\boldsymbol{L}}
\newcommand{\SV}{\boldsymbol{S}}
\newcommand{\av}{\boldsymbol{a}}
\newcommand{\qv}{\boldsymbol{q}}
\newcommand{\QV}{\boldsymbol{Q}}
\newcommand{\ev}{\boldsymbol{e}}
\newcommand{\uv}{\boldsymbol{u}}
\newcommand{\KV}{\boldsymbol{K}}
\newcommand{\ro}{\boldsymbol{\rho}}
\newcommand{\si}{\boldsymbol{\sigma}}
\newcommand{\thv}{\boldsymbol{\theta}}
\newcommand{\bv}{\boldsymbol{b}}
\newcommand{\JV}{\boldsymbol{J}}
\newcommand{\nv}{\boldsymbol{n}}
\newcommand{\lv}{\boldsymbol{l}}
\newcommand{\om}{\boldsymbol{\omega}}
\newcommand{\Om}{\boldsymbol{\Omega}}
\newcommand{\Piv}{\boldsymbol{\Pi}}
\newcommand{\UV}{\boldsymbol{U}}
\newcommand{\iv}{\boldsymbol{i}}
\newcommand{\nuv}{\boldsymbol{\nu}}
\newcommand{\muv}{\boldsymbol{\mu}}
\newcommand{\lm}{\boldsymbol{\lambda}}
\newcommand{\Lm}{\boldsymbol{\Lambda}}
\newcommand{\opsi}{\overline{\psi}}
\renewcommand{\tan}{\textrm{tg}}
\renewcommand{\cot}{\textrm{ctg}}
\renewcommand{\sinh}{\textrm{sh}}
\renewcommand{\cosh}{\textrm{ch}}
\renewcommand{\tanh}{\textrm{th}}
\renewcommand{\coth}{\textrm{cth}}

\begin{abstract}
We take advantage of the Shiromizu et al. covariant formalism to find out the brane properties originating from the five dimensional bulk spacetime. Making a different choice for the conformal factor $e^{-2b(z)}$ compared to Estrada \cite{ME}, we reach a new solution with a lot of interesting properties, where $b(z) = ln(1/\sqrt{cosh2\mu z})$. The non-local tensor $E_{ab}$ rooted from the 5-dimensional Riemann tensor gives an anisotropic stress energy tensor on the brane with positive energy density and negative radial pressure. The BH on the brane looks like a black string in the 5-dimensional space, with no singularities of the curvature invariants.

 \end{abstract}
 
\section{Introduction}
Since the pioneering works of Randall and Sundrum (RS) \cite{RS1, RS2}, several papers developed the subject of spacetimes embedded in 5-dimensional geometries \cite{GL, MP, GH, HM, HC4, DM, TW, CHR, ER}. The approach has been highlighted by String theory which contains gravity and requires more than four dimensions. In fact, the first successful statistical counting of black hole entropy in string theory was performed for a five-dimensional black hole \cite{SV}.

In the first model, RS assumed that our universe is a negative tension domain wall separated from a positive tension wall by a slab of anti-de Sitter (AdS), a large extra dimension not being necessary, so that the hierarchy problem is solved by the special properties of AdS. However, there are troubles with the negative tension of one of the domain walls. In their second model \cite{RS2}, one considers we live on a positive tension brane inside anti-de Sitter space, the spacetime being reflection symmetric. A black hole (BH) created by gravitational collapse will have a horizon that extends into the dimensions transverse to the brane: it will be a higher dimensional object \cite{CHR}. 

Gregory and Laflamme \cite{GL} showed that the black strings (BS) (i.e. 5-dimensional objects with an event horizon having topology $S^{2} \times S^{1}$ - according to \cite{HM}), are unstable to linearized perturbations. However, Horowitz \cite{GH} proved that an event horizon cannot pinch off in finite time, or it cannot shrink to zero size in finite
affine parameter. His basic argument was that the scalar expansion of the null geodesic generators of the horizon cannot become negative.

Chamblin, Hawking and Reall \cite{CHR} stated that what looks like a BH on the brane is actually a BS in the higher dimensional space. They also showed that the BS is unstable only near the AdS horizon (the Gregory-Laflamme instability) but it is stable far from it. Then they describe a BH localized in the fifth dimension - the Schwarzschild-AdS solution. Maartens \cite{RM2} reinforced the Chamblin-Hawking-Reall analysis, showing that the singularity r = 0 is a line along the z-axis, so that the Schwarzschild-AdS bulk metric describes a ''black string''. The 5-dimensional horizon is the surface $g_{tt} = 0$ in the bulk. The bulk horizon is a sphere of Schwarzschild radius on each z = constant surface, so that it has a cylindrical shape in the z-direction.

In the brane-world scenario, one considers that the standard matter fields are confined to the 4-dimensional spacetime (the brane) whereas gravity propagates in the full spacetime (the bulk) \cite{RM1, SMS, HC1}. Shiromizu et al. formulated covariant equations that describe both the 5-dimensional gravity in the bulk and the 4-dimensional gravity on the brane. They found that a positive tension brane has the correct sign of gravity and their equations become the conventional Einstein equations in the low energy limit. 

Dadhich et al. \cite{DMPR} (see also \cite{GM}) studied models of spherically-symmetric stars and black holes (BH) localized on a three-brane in 5-dimensional gravity in the Randall-Sundrum (RS) scenario \cite{RS2}. Dadhich et al. stated that the Reissner-Nordstrom geometry is an exact solution of the effective Einstein equations on the brane, a BH with a tidal ''charge'' arising via gravitational effects from the fifth dimension. The solution satisfies a closed system of equations on the brane, describing a strong-gravity regime. 

If our universe is to be regarded as a domain wall in five-dimensional AdS spacetime, how could we check that experimentally? Perhaps the spacetime is effectively five dimensional below the Grand Unified scale \cite{LOW} and our universe arises as a domain wall or brane, at its boundary \cite{RG2}. There is the possibility of observing fundamental
spin-2 particles at LHC (as pointed out in \cite{HDD, AHDD}). The authors of \cite{HDD, AHDD} proposed a new framework for solving the hierarchy problem: the gravitational and gauge interactions become united at the weak scale, which they have taken as the fundamental short distance scale in nature.  This picture leads to a number of striking signals for accelerator and laboratory experiments (for instance, the creation of small BHs during particle collisions at the LHC \cite{PK}). For the case of n = 2 new dimensions, the sub-millimeter measurements of gravity may observe the transition from $1/r^{2} \rightarrow 1/r^{4}$ Newtonian gravitation. For any number of new dimensions, the LHC could observe strong quantum gravitational interactions. However, to the best knowledge of the author, strong quantum gravitational effects have not been observed at LHC by now.

More recently, Estrada \cite{ME} provides a new five dimensional solution by embedding a four dimensional regular BH (the Hayward metric) into a compact extra dimension, with an asymptotically flat four dimensional geometry at infinity of the radial coordinate and a de Sitter core instead of a central singularity.

Motivated by Estrada's paper, we make use of a different regular Schwarzschild-type metric \cite{HC2, HC3} with an exponential factor that makes finite all physical quantities (including invariants) at the origin and at infinity. In addition, our constraint equation of the conformal factor of the 4-dimensional line-element on the brane is different from that used by Estrada \cite{ME}, such that the conformal factor is an even function of the fifth dimension. Section 2 presents the 5-dimensional geometry in terms of a conformal factor $b(z)$ in front of a regular Schwarzschild-like metric. The main Section 3 investigates the properties of the anisotropic energy-momentum tensor from the bulk and the induced metric on the brane by means of the Shiromizu et al. covariant recipe. The components of the tidal nonlocal tensor $E_{ab}$ from the bulk are also computed. The conclusions are resumed in Section 4.
  We also intend to investigate the stability problem of the BS in a future work.

\section{The 5-dimensional regular geometry}
The Einstein gravitational equation in 5D looks like
    \begin{equation}
		^{5}G_{ab} \equiv ^{5}R_{ab} -\frac{1}{2}g_{ab} ~^{5}R^{c}_{~c} = 8\pi G_{5}~ ^{5}T_{ab},
 \label{2.1}
 \end{equation}
where $a, b = 0, 1, 2, 3, 5$ and $G_{5}$ is the 5-dimensional Newton constant (we take, for convenience, $8\pi G_{5} = 1 = c$, where $c$ is the velocity of light. Our regular 5-dimensional geometry is considered as 
  \begin{equation}
	 ds^{2} = e^{-2b(z)}\left[-\left(1 - \frac{2m}{r} e^{-\frac{k}{r}}\right) dt^{2} + \frac{1}{1 - \frac{2m}{r} e^{-\frac{k}{r}}} dr^{2} + r^{2} d \Omega^{2}\right] + dz^{2},     
 \label{2.2}
 \end{equation} 
where $b(z)$ is a function of the fifth coordinate $z$, $k$ is a positive constant, $m$ is the BH mass parameter, $d \Omega^{2}$ stands for the metric on the unit two-sphere and coordinates are $(t, r, \theta, \phi, z)$. The expression within the square parantheses represents the regularized Schwarzschild geometry in 4-dimensions \cite{HC2, HC3}. The spacetime (2.2) is endowed with a horizon, depending on the value of $k$ w.r.t. $2m/e$, with $lne = 1$. If $k = 2m/e$ (the extremal case), the horizon is located at $r = 2m/e$, where $g_{tt} = 0$. For $k>2m/e$, the metric coefficient $f(r) = 1 - \frac{2m}{r} e^{-\frac{k}{r}}$ is strictly positive and there is no any horizon. In contrast, when $k<2m/e$ there are two horizons and $f(r)$ becomes negative between them. 

 It is worth noting that any $z = const.$ three brane gives us a geometry conformal to a 4-dimensional regular Schwarzschild geometry. The conformal factor $b(z)$ gives the scale of the line element. One also observes that the brane 4-dimensional metric is regular both at the origin $r = 0$ and when $r \rightarrow \infty$. In addition, the solution looks like a black string viewed from the five dimensional spacetime.

For the metric (2.2), Einstein's equations (2.1) are given by
  \begin{equation}
	\begin{split} 
	^{5}G^{t}_{~t} = -\frac{2mk~e^{-\frac{k}{r}}} {r^{4} e^{-2b(z)}} + 6b'^{2} - 3b'' = - \rho,~~~ ^{5}G^{r}_{~r} = ^{5}G^{t}_{~t}= p_{r}, \\
	^{5}G^{\theta}_{~\theta} = \frac{2mk~e^{-\frac{k}{r}}} {r^{4}e^{-2b(z)} } \left(1 - \frac{k}{2r}\right) + 6b'^{2} - 3b'' = p_{\theta},\\ ^{5}G^{\phi}_{~\phi} = ^{5}G^{\theta}_{~\theta}= p_{\phi}, ~~~^{5}G^{z}_{~z} = -\frac{mk^{2}e^{-\frac{k}{r}}} {r^{5}e^{-2b(z)}} + 6b'^2 = p_{z},
 \label{2.3}
\end{split} 
 \end{equation} 
where $b' = db/dz$ and $\rho, p_{r}, p_{\theta}, p_{\phi}, p_{z}$ are, respectively, the energy density, radial pressure and transversal pressures. 
 
 To simplify his system of differential equations, Estrada \cite{ME} assumed that the scale factor $b(z) (A(z)$ in his notations) has a constant derivative $b' = db/dz$ (related with his cosmological constant) and so he obtained $b''(z) = 0$. We choose a different option and take
  \begin{equation}
	b'' - 2b'^{2} = -2\mu^{2},
 \label{2.4}
 \end{equation} 
where $\mu$ is a positive constant and $b''\neq 0$. It is easy to check that 
  \begin{equation}
	b(z) = - \frac{1}{2} ln(cosh2\mu z)
 \label{2.5}
 \end{equation} 
 is a solution of the differential equation (2.4). The scale factor becomes $e^{-2b(z)} = cosh 2\mu z $. Note that $b(z) \leq 0$, with $b(0) = 0$. It becomes infinite when $z \rightarrow \infty$. As Kanti et al. have noticed \cite{KOT}, doubts appear on localization of gravity in that case. A solution to localize gravity would be to take z a compact coordinate \cite{KOT}, an option we admit hereafter.

Moreover, the geometry (2.2) does not depend on the sign of $z$ because $cosh2\mu z$ is an even function and it is also continuous and differentiable. Our choice for $b(z)$ is justified by the simple form of the Einstein equations (2.3). 

With the above choice of $b(z)$, we intend now to find how $b(z)$ affects the acceleration of a static observer in the spacetime (2.2). The velocity vector field for a static observer appears as 
\begin{equation}
	u^{a} = \left(\frac{1}{\sqrt{(1 - \frac{2m}{r} e^{-\frac{k}{r}})cosh2\mu z}}, 0, 0, 0, 0\right).
 \label{2.6} 
 \end{equation} 
The nonzero components of the covariant acceleration $a^{b} = u^{a}\nabla_{a}u^{b}$  are given by
  \begin{equation}
	a^{r} = \frac{m(1 - \frac{k}{r})} {r^{2} cosh 2\mu z} e^{-\frac{k}{r}}, ~~~a^{z} = \mu ~tanh 2\mu z .
 \label{2.7}
 \end{equation} 
   $a^{b}$ is the acceleration necessary for maintaining the observer static position. It is worth noting that $a^{r}$ changes its sign at $r = k$. It also vanishes for $r \rightarrow 0$ or $r \rightarrow \infty$, at constant $z$. In addition, $a^{z}$ is an acceleration in the off-brane direction \cite{RM1}, rooted from the nonlocal gravitational field from the bulk. It depends only on z-variable and vanishes at $z = 0$. 
	
	A test particle located at, say,  $r > k = 2m/e$ will have $a^{r}>0$; namely, the particle will be attracted by the source $m$. In contrast, if the particle is located at $r < k = 2m/e$, it will be rejected towards the horizon $r = 2m/e$. For $k<2m/e$, we have two horizons \cite{HC2}, obtained from $1 - \frac{2m}{r_{H}} e^{-\frac{k}{r_{H}}} = 0$. In a future work we intend to examine the surface gravity and pursue the thermodynamic behaviour of the BH.  
			
	As far as the curvature invariants are concerned, for the scalar curvature we get
  \begin{equation}
	^{5}R^{a}_{~a} = \frac{2mk^{2}e^{-\frac{k}{r}}} {r^{5} cosh 2\mu z} -16\mu^{2} - 4\mu^{2} tanh^{2} 2\mu z ,
 \label{2.8}
 \end{equation} 
that is finite everywhere, for any values of $r$ and $z$. It acquires the constant value $-16\mu^{2}$ in the limit  $r \rightarrow 0$ and $z = 0$. The Kretschmann scalar $K \equiv R^{abcd}R_{abcd}$ is given by
  \begin{equation}
	\begin{split}
	K = \frac{48m^{2}}{r^{6}cosh^{2} 2\mu z} \left(1 - \frac{2k}{r} + \frac{2k^{2}}{r^{2}} - \frac{2k^{3}}{3r^{3}} + \frac{k^{4}}{12r^{4}}\right) e^{-\frac{2k}{r}} \\
	-\frac{8mk^{2}\mu^{2}sinh^{2}2\mu z}{r^{5}cosh^{3}2\mu z} e^{-\frac{k}{r}} + \frac{8\mu^{4}}{cosh^{4}2\mu z} (5cosh^{4}2\mu z -2cosh^{2}2\mu z +5).
 \label{2.9}
\end{split}
 \end{equation} 
It is also finite for any $r$ or $z$ (in contrast, one observes that the Kretschmann scalar of Chamblin et al. \cite{CHR} is divergent when $z \rightarrow \infty$ or $r \rightarrow 0$).

  With $b(z)$ from (2.5), it is clear that the conformal factor is continuous and differentiable. The regular BH on the brane appears as a black string (BS) in the bulk. There is no any horizon on the bulk but only that one on the brane, calculated from $f(r) = 0$~\footnote{If $k>2m/e$, the  equation $f(r) = 0$ has no real roots and, without a horizon on the brane, there is no any BS in the bulk.}. As in Maartens \cite{RM2} example, our BS has a cylindrical shape in the z-direction (see also \cite{ME}) but it is a sphere on the brane, with the bulk horizon radius obtained from $g_{tt} = 0$. 

\section{The anisotropic stress tensor}
The spacetime (2.2) with the choice (2.5) is generated by the following energy-momentum tensor
  \begin{equation}
	\begin{split}
	^{5}T^{t}_{~t} = ^{5}T^{r}_{~r} = -\frac{2mk~e^{-\frac{k}{r}}} {r^{4} cosh 2\mu z} + 6\mu^{2},~~~  \\
	^{5}T^{\theta}_{~\theta} = ^{5}T^{\phi}_{~\phi} =  \frac{2mk~e^{-\frac{k}{r}}} {r^{4} cosh 2\mu z} \left(1 - \frac{k}{2r}\right) + 6\mu^{2}\\
	^{5}T^{z}_{~z} = -\frac{mk^{2}e^{-\frac{k}{r}}} {r^{5} cosh 2\mu z} + 6\mu^{2} tanh^{2} 2\mu z .
 \label{3.1}
\end{split}
 \end{equation} 
Let us noting that the (minus) energy density $\rho$ and the pressures $p_{r}, p_{\theta}, p_{\phi}$ (but not $p_{z}$) get the constant value $6\mu^{2}$ if $r$ tends to zero or infinity, irrespective of the value of $z$. In addition, with increasing $z$ and for any $r$, all the components of the stress tensor tends to the same value $6\mu^{2}$.

 We are now studying the impact of the 5-dimensional gravitational field (expressed by the Riemann/Weyl tensor) upon the three-brane properties. By means of the covariant decomposition developed by Shiromizu et al.\cite{SMS} we write down the effective gravitational equations on the brane. The stress tensor $T_{ab}$ on the brane is induced by the 5-dimensional bulk via the tidal nonlocal tensor $E_{ab}$, expressed in terms of the 5D Riemann tensor. 

Let us take for the time being a general bulk spacetime with five dimensions. Our 4-dimensional world is described by a three-brane embedded in 5-dimensional space. Let $n^{a}$ be the spacelike unit vector, normal to the brane and $h_{ab} = g_{ab} - n_{a}n_{b}$, the induced metric on the brane ($g_{ab}$ is the full 5-dimensional metric). Shiromizu et al. \cite{SMS} (see also \cite{HC1}) have shown that, from the Gauss equations relating the Riemann tensors in 5- and 4-dimensions and the Codazzi equations for the variation of the extrinsic curvature, one readily obtains
  \begin{equation}
  \begin{split}
  G_{ab} = (^{5}R_{cd} - \frac{1}{2}g_{cd}~ ^{5}R)h^{c}_{a}h^{d}_{b} + ^{5}R_{cd}n^{c}n^{d}h_{ab} + K_{ab} K - K^{c}_{a}K_{bc}\\ - \frac{1}{2}h_{ab}(K^{2} - K^{cd} K_{cd}) - E_{ab}
  \end{split}
 \label{3.2}
 \end{equation} 
where $^{5}R_{ab}$ is the 5-dimensional Ricci tensor, $K = K^{a}_{~a}$ is the trace of the extrinsic curvature $K_{ab} = h^{c}_{a}h^{d}_{b}\nabla_{c}n_{d}$ and 
 \begin{equation}
 E_{ab} =  ^{5}R^{c}_{~def}n_{c}n^{e}h^{d}_{~a}h^{f}_{~b}
 \label{3.3}
 \end{equation}
is the tensor related to the tidal effects on the brane, coming from the bulk. 

We wish now to apply the Shiromizu et al. formalism for the 5-dimensional metric (2.2). We choose, for convenience, the brane to be located on the hypersurface $z = z_{0}$, and the normal to the brane is $n^{a} = (0, 0, 0, 0, 1)$. For the components of the extrinsic curvature tensor one obtains
 \begin{equation}
\begin{split}
K_{tt} = -\mu (1 - \frac{2m}{r} e^{-\frac{k}{r}}) sinh 2\mu z,~~~K_{rr} = \frac{\mu sinh 2\mu z}{1 - \frac{2m}{r} e^{-\frac{k}{r}}}\\
K_{\theta \theta} = \mu r^{2} sinh 2\mu z,~~~K_{\phi \phi} = K_{\theta \theta} sin^{2} \theta,~~~K = 4\mu~ tanh2\mu z ,
\end{split}
 \label{3.4}
 \end{equation} 
evaluated at $z = z_{0}$. We have also
  \begin{equation}
	^{5}R_{cd}n^{c}n^{d} = -4\mu^{2} \frac{1+cosh^{2}2\mu z}{cosh^{2}2\mu z}
 \label{3.5}
 \end{equation} 
 Eq.3.2 yields now
  \begin{equation}
	\begin{split}
	E_{tt} = \mu^{2} (1 - \frac{2m}{r} e^{-\frac{k}{r}}) \frac{1+cosh^{2}2\mu z}{cosh2\mu z},~~~E_{rr} = -\frac{\mu^{2}}{1 - \frac{2m}{r} e^{-\frac{k}{r}}} \frac{1+cosh^{2}2\mu z}{cosh2\mu z}\\ E_{\theta \theta} = - \mu^{2} r^{2} \frac{1+cosh^{2}2\mu z}{cosh2\mu z},~~~  E_{\phi \phi} = E_{\theta \theta} sin^{2}\theta.
\end{split}	
 \label{3.6}
 \end{equation} 

As is well known, in the Newtonian gravitation the tidal effects come from terms containing $m/r^{3}$, which produces a shear in a timelike or null geodesic congruence. From (3.3) it is clear that $E_{ab}$ is directly related to the Riemann tensor and $m/r^{3}$ is one of the basic components of the Riemann tensor (for example, for the Schwarzschild spacetime). Moreover, one notes that, in the mixed components form, we have $E^{t}_{~t} = E^{r}_{~r} = E^{\theta}_{~\theta} = E^{~\phi}_{\phi} = -\mu^{2} - \mu^{2}/cosh^{2}2\mu z$, with a clear dependence of $E^{a}_{~b}$ on the constant $\mu$ and the fifth coordinate $z$ only. 

Once we have computed all the quantities entering (3.2), we are now in a position to find the stress tensor induced on the brane. One obtains
  \begin{equation}
	\begin{split}
	T^{t}_{~t} = ^{5}T^{t}_{~t} - 6\mu^{2} = -\frac{2mk~e^{-\frac{k}{r}}} {r^{4} cosh 2\mu z} = T^{r}_{~r}\\ T^{\theta}_{~\theta} = ^{5}T^{\theta}_{~\theta} - 6\mu^{2} = \frac{2mk~e^{-\frac{k}{r}}} {r^{4} cosh 2\mu z} \left(1 - \frac{k}{2r}\right) = T^{\phi}_{~\phi},
\end{split}	
 \label{3.7}
 \end{equation} 
evaluated at $z = z_{0}$. As an overall observation, we notice that the contribution of the fifth dimension on the induced stresses on the brane is the factor $1/cosh~2\mu z$.

 We finally stress that the BH on the brane looks like a black string in the 5-dimensional space, with no singularities of the curvature invariants.

\section{Conclusions}
A new restriction imposed on the conformal factor depending on the fifth coordinate $z$ gives a lot of interesting properties for the physical quantities both on the bulk and on the three brane embedded in it. The stress tensor in the bulk depends on $z$ and the radial coordinate $r$, being finite for any values of them. It depends also on the free parameters $m$-the Schwarzschild mass and $k, \mu$- positive constants. We computed the energy-momentum tensor induced on the brane by the nonlocal tensor $E_{ab}$ generated by the 5D Riemann tensor. 
The BH on the brane looks like a black string in the 5-dimensional space, with no singularities of the curvature invariants.\\

\textbf{Acknowledgements}

I am grateful to the anonymous referee for useful suggestions which considerably improved the quality of the manuscript.

\end{document}